\documentclass{llncs}

\usepackage{oz}
\usepackage{times}
\usepackage{url}
\usepackage{alltt}
\usepackage{graphicx}

\newlength{\zedwidth}
\def\plus{{^+}}
\def\star{{^*}}

\begin {document}

\zedtab=1.5em
\zedindent=1em

\def\fuzz{{\large\it f}{\sc uzz}}

\def\emquote{\vspace{-.8ex}\begin{quote}\small\em}
\def\endemquote{\end{quote}\vspace{-.8ex}}

\def\zsidebyside{\begin{sidebyside}}
\def\endzsidebyside{\vspace{-2ex}\end{sidebyside}}

\mainmatter

\title{Provably Correct Systems: \\
Community, connections, and citations}

\author{Jonathan P.\ Bowen}
\institute{
Centre for Software Engineering \\
Birmingham City University, UK \\
\email{jonathan.bowen@bcu.ac.uk} \\
\url{www.jpbowen.com}
}

\date{}

\maketitle

\begin{abstract}
The original European ESPRIT ProCoS I and II projects on {\em
Provably Correct Systems} took place around a quarter of a
century ago. Since then the legacy of the initiative has spawned
many researchers with careers in formal methods. One of the
leaders on the ProCoS projects was Ernst-R\"udiger Olderog. This
paper charts the influence of the ProCoS projects and the
subsequent ProCoS-WG Working Group, using Prof.\ Dr Olderog as
an example. The community of researchers surrounding an
initiative such as ProCoS is considered in the context of the
social science concept of a Community of Practice (CoP) and the
collaborations undertaken through coauthorship of and citations
to publications. Consideration of citation metrics is also
included.
\end{abstract}

%

\section{Background}
\label{background}

Historically, the creation of scientific knowledge has relied on
collaborative efforts by successive generations through the
centuries \cite{Van91}. Scientific advances are gradually
developed by a community of researchers over time (e.g., the
abstract algebra of the French mathematician \'Evariste Galois
leading to Galois theory and group theory \cite{Bow15}). A
scientific theory can be modelled as a mathematical graph of
questions posed by scientists (represented by the vertices of
the graph) and the corresponding answers (modelled by arcs
connecting the vertices in the graph) \cite{San96}. The answers
to questions lead to further questions and so the process
continues, potentially ad infinitum. In general, mathematical
logic underlies the valid reasoning that is required for
worthwhile development of scientific theories and knowledge
\cite{Har72}.

In recent years, the speed of transmission and the quantity of
knowledge available has accelerated dramatically, especially
with improvements in the Internet and specifically the
increasing use of the World Wide Web \cite{Ber00}. Whereas
previously academic papers were published on paper in journals,
conference proceedings, technical reports, books, etc., now all
these means of communication can and often are done largely
electronically online. The plethora of information has also
become indexed more and more effectively, especially with the
advent of the PageRank algorithm as used by Google \cite{Lan06}.

In this paper, we use the European ProCoS (``Provably Correct
Systems'') initiative of the 1990s
\cite{jpb:Bjorner89,procos2:LR94} as an example of a
foundational community of academic researchers working in
various areas towards a common aim. We consider the related
issue of the production of publications and their citations as
an important aspect of scholarly activity. We model some aspects
of this formally using the Z notation \cite{Bow01,Spi01} to help
in disambiguating some of the concepts that are often left
somewhat nebulous in social science (e.g., with respect to a
Community of Practice \cite{Wen98,Wen02}).

Section~\ref{ProCoS} introduces the European collaborative
ProCoS projects and the subsequent Working Group of the 1990s.
In Section~\ref{community}, we present an example ProCoS
researcher and their relationship with other researchers through
coauthorship and citations, with visualizations of these
relationships. The Section formalizes the relationship of
researchers in an academic community such as that generated by
ProCoS and extends this to cover a Community of Practice.
Section~\ref{metrics} considers some of the citation metrics
that are available for measuring a researcher's influence,
including their shortcomings, using publication corpuses that
are now available online. Finally Section~\ref{conclusion}
provides a conclusion and some possible future directions.

\section{The ProCoS Community}
\label{ProCoS}

In this section, we consider the development of the ProCoS
initiative and the community that it has created.
The seeds of the ProCoS projects on ``Provably Correct Systems''
took place in the 1980s \cite{jpb:Bjorner89,procos2:LR94}, coming out
of the formal methods community \cite{Boc10,Bow14a}. The CLInc
Verified Stack initiative of Computational Logic Inc.\ in the
USA \cite{jpb:M*89,procos2:Young94}, using the Boyer-Moore Nqthm
theorem proving to verify a linked set of hardware, kernel and
software in a unified framework, was an inspiration for the
initial ProCoS project. Whereas CLInc was a closely connected
set of mechanically proved layers, ProCoS concentrated more on
possible formal approaches to the issues of verifying a complete
system at more levels from requirements, specification, design,
and compilation, using a diverse set of partners around Europe
with different backgrounds, expertise, and interests, but with a
common overall goal. A ProCoS ``tower'' with appropriate
formalisms and approaches was proposed to investigate proving a
system correct in a linked way at the various levels of
abstraction. The approach was based around the Occam parallel
programming language and Transputer microprocessor architecture.
A gas burner was used as a motivating example for much of the
work.

The first ProCoS project was for $2\frac{1}{2}$ years
(1989--1991) with seven academic partners \cite{jpb:Bjorner89}.
The subsequent ProCoS II project (1992--1995) involved a more
focused set of four academic partners \cite{jpb:Bowen*93}.
Subsequently a ProCoS-WG Working Group of 25 partners allowed a
more diverse set of researchers to engage in the ProCoS
approach, including industrial partners \cite{jpb:Bowen*94}.

The ProCoS projects worked on various aspects of formal system
development at different related levels of abstraction,
including program compilation from an Occam-based programming
language to a Transputer-based instruction set
\cite{procos2:BFOR93,procos2:FTRTFToverview,procos2:LR94}. A gas
burner was used extensively as a case study and this helped to
inspire the development of Duration Calculus for succinctly
formalized real-time requirements \cite{procos2:ZHR91}. A novel
provably correct compiling specification approach was also
developed using a compiling relation for the various constructs
in the language that could be proved using algebraic laws
\cite{jpb:HHBP90}. This was later extended to a larger language
including recursion \cite{jpb:HB92,He95}. The project used
algebraic and operational semantics in its various approaches.
The relationship between these and also denotational semantics
was later demonstrated more universally in the {\em Unified
Theories of Programming} (UTP) approach \cite{procos2:HH98}.

A Community of Practice (CoP) \cite{Wen98,Wen02} is widely
accepted social science approach used as a framework in the
study of the community-based process of producing a particular Body
of Knowledge (BoK) \cite{Bow11}. An example of a CoP is that
generated by the ProCoS initiative in the area of provably
correct systems \cite{procos2:BFOR93,procos2:FTRTFToverview}, as
discussed later in this paper. The important elements of a CoP
include a domain of common interest (e.g., provably correct
systems), a community willing to engage with each other (e.g.,
members of the ProCoS projects and Working Group), and
exploration of new knowledge to improve practice (e.g., Duration
Calculus \cite{procos2:ZHR91} and later UTP \cite{procos2:HH98}).


\section{A Community around a Researcher}
\label{community}

\begin{figure}[bth]
\begin{center}
\includegraphics[width=\textwidth]{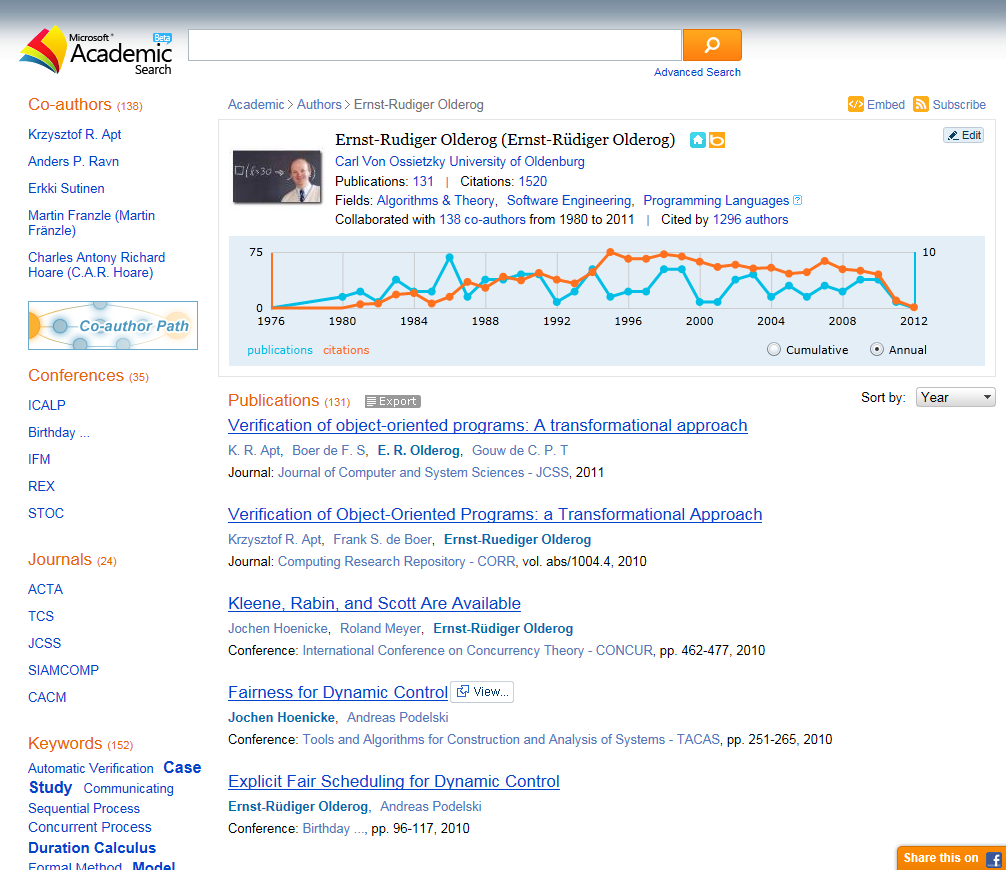}
\end{center}
\caption{Publication and citation statistics
for Ernst-R\"udiger Olderog on Academic Search.\label{ERO-home}}
\end{figure}

Here we use the German computer scientist and one of the
original leaders on the ProCoS project, Ernst-R\"udiger Olderog
\cite{procos2:Olderog91,procos2:Olderog92,procos2:Olderog94}
of the University of Oldenburg, as an example of a leading
member of a community of researchers. The visualization
capabilities of the online Microsoft Academic Search facility
(\url{http://academic.research.microsoft.com}) are used to
illustrate this. For example, see Figure~\ref{ERO-home} for
E.-R.\ Olderog's home page on Academic Search. The Academic
Search facilities include graphical presentation of direct
relationships of collaborators as coauthors of publications,
direct citations of other researchers to an individual's
publications, and indirect connections between any two authors
through intermediate coauthors.

Academic Search also lists the coauthors, conferences and
journals for each author in reverse order of publication count
and the main keywords associated with the publications of an
author (see Figure~\ref{ERO-home}). For example, three out of
the top five coauthors of E.-R.\ Olderog were associated with
the ProCoS project. In addition, he is particularly active in
the {\em International Colloquium on Automata, Languages, and
Programming} (ICALP), the {\em Integrated Formal Methods} (IFM)
conferences, and the {\em Acta Informatica} and {\em Theoretical
Computer Science} journals (again, see Figure~\ref{ERO-home}),
Important keywords include ``Duration Calculus'', a direct
result (and unpredicted) of the ProCoS project.


The links between coauthors and citing authors form mathematical
graphs \cite{Bow12b}. These can be modelled using relations. The
Z notation \cite{Hen03,Spi01} is a convenient notation to
present these formally, as previously demonstrated in
\cite{Bow13b}. Here we concentrate on authors rather than
publications and the paths of coauthors that connect
researchers. In particular, we augment this model to
consider the collaborative distance between an arbitrary pair of
authors in terms of coauthorship. We model all the possible
paths between such authors as a set of sequences of authors
where the two authors under consideration are the first and last
author in each of the sequences. The two authors also do not
occur within these sequences and authors are not repeated in the
sequences either.

We use the concept of graphs in our mathematical modelling.
A general graph can be modelled as a relation in Z, using a
generic constant on any set $X$:

\begin{gendef}[X]
graph : X \rel X \\
\end{gendef}

We can refine a general graph and consider a model for an
undirected graph in Z:

\begin{gendef}[X]
ugraph : \power graph \\
\where
ugraph = ugraph\inv
\also
ugraph\cap \id X = \emptyset
\end{gendef}

\noindent
Here all nodes (authors) are connected in both directions (as
coauthors) and also a node cannot be connected to itself (i.e.,
an author cannot be a coauthor with themselves). In the above
definition, ``$\inv$'' indicates the inverse of a relation and
``$\id$'' produces the identity relation from a set.

%
%
%
%

Academic communities consist of people that have authored
publications. In Z, this can be modelled as a given set:

\def\PEOPLE{PEOPLE}

\begin{zed}
[ \PEOPLE ]
\end{zed}

In an academic community of researchers for a particular area,
there is often a main key researcher leading the field's
publications. Then there is a wider number of researchers that
have published papers in the field. Typically published works
have a number of coauthors. Published authors may be related to
other authors transitively through coauthorship. Authors may
also be cited by other published authors, even if not related
through coauthorship. These relationships can be modelled
formally using graphs:

\begin{schema}{Researchers}
main : \PEOPLE \\
published : \finset_1 \PEOPLE \\
coauthors, related, citing\_authors : \PEOPLE \rel \PEOPLE
\where
main \in published \\
coauthors \subseteq ugraph[published] \\
related \subseteq ugraph[published] \\
related = coauthors\plus \\
citing\_authors \subseteq graph[published]
\end{schema}

\noindent
Note that ``$\finset_1$'' indicates a finite non-empty set and
``$\plus$'' indicated irreflexive transitive closure above.

\begin{figure}[bth]
\begin{center}
\includegraphics[width=\textwidth]{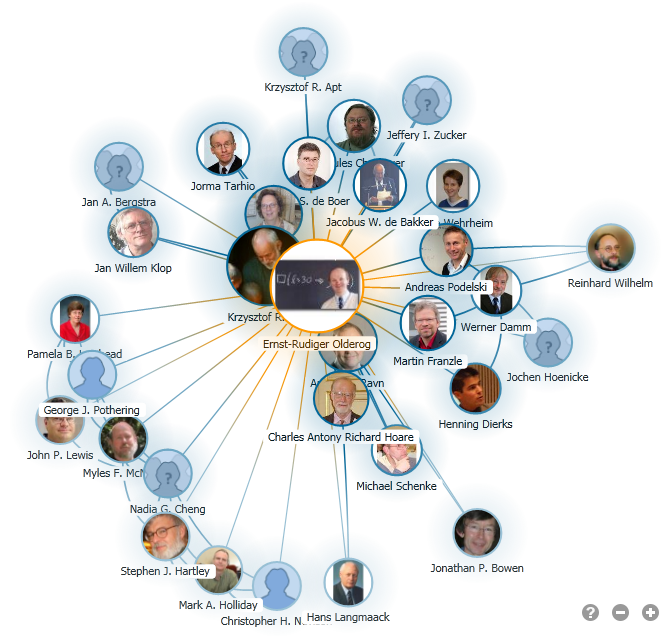}
\end{center}
\caption{Primary coauthors of
Ernst-R\"udiger Olderog on Academic Search.\label{ERO-coauthors}}
\end{figure}

\begin{figure}[bth]
\begin{center}
\includegraphics[width=\textwidth]{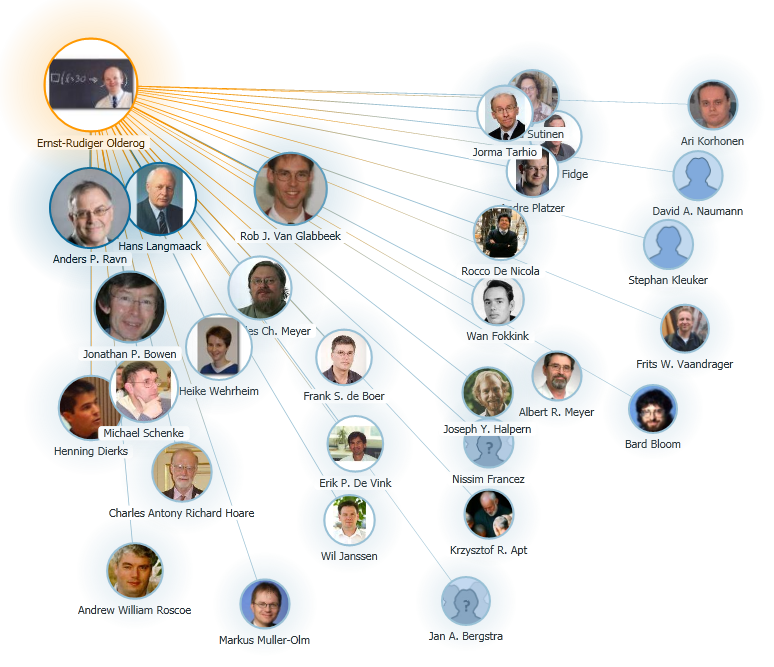}
\end{center}
\caption{Primary citing authors
for Ernst-R\"udiger Olderog on Academic Search.\label{ERO-citations}}
\end{figure}

The Academic Search facility enables graphical visualization of
the coauthors (e.g., see Figure~\ref{ERO-coauthors}) and citing
authors (e.g., see Figure~\ref{ERO-citations}) for any
particular $author$ in its database. Figure~\ref{ERO-coauthors}
provides a pictorial view of a subset of the relation
$\{author\} \dres related \rres coauthors \limg \{author\}
\rimg$ (where ``$\dres$'' indicates domain restriction of a
relation, ``$\rres$'' indicates range restriction of a relation,
and ``$\limg \ldots \rimg$'' indicates a relational image of a
subset of the domain) for a specific $author$ (in this case
Ernst-R\"udiger Olderog) at the centre.  Connections between
coauthors who have themselves written publications together can
be shown as well, in addition to coauthorship with the main
author under consideration. This results in groupings of
coauthors that are interconnected in a way than can be seen
visually very quickly.  For example, in this case all the
coauthors associated with the ProCoS project are in the lower
right-hand quadrant, including the author of this paper.

Figure~\ref{ERO-citations} gives a pictorial view
of a subset of the relation $\{author\} \dres citing\_authors$,
again for a specific $author$ located at the top left position
in the diagram. Citations from authors involved with the ProCoS
project are largely grouped on the left-hand side of the
diagram, during Olderog's early career. Later citations are to
the right.

Next we consider paths between pairs of nodes (authors):

\begin{gendef}[X]
path : (X\cross X) \rel \iseq X
\where
\forall x_1, x_2 : X; s : \iseq X \mid \#s>1 \spot {} \\
\t1 (x_1,x_2)\mapsto s \in path \iff {} \\
\t2 head~s = x_1 \land {} \\
\t2 last~s = x_2 \land {} \\
\t2 (\forall n:\nat_1 \mid n<\#s \spot (s~n, s(succ~n)) \in graph)
\end{gendef}

\noindent
The paths are modelled as injective sequences (``$\iseq$'') of
length more than one, where the first and last entries in the
sequences are the two nodes under consideration and all adjacent
pairs in the sequence are directly connected in the graph.
Because the sequences are injective, no nodes are repeated in
these sequences. This means that the pair of nodes under
consideration are always two different nodes.

The collaborative distance of two authors can be of particular
interest. Two authors may be connected in many different ways by
sequences of coauthors or even in no way whatsoever (effectively
an infinite collaborative distance). The shortest (minimum)
connection between two different authors is of special interest.

\begin{gendef}[X]
dist : X\cross X \pfun \nat_1
\where
\forall x_1, x_2 : X \mid (x_1, x_2) \in \dom path \spot {} \\
\t1 dist(x_1,x_2) =
      min ( \# \limg path \limg \{ (x_1,x_2) \} \rimg \rimg )
\end{gendef}

In recent years, the ``Erd\H{o}s number'' (i.e., the
collaborative distance from Erd\H{o}s) has become a metric for
involvement in mathematical and even computer science research
\cite{Bow12b}. Paul Erd\H{o}s, a very collaborative 20th century
mathematician, is considered to have an Erd\H{o}s number of 0.
His direct coauthors (511 of them) have an Erd\H{o}s number of
1. Other authors can be assigned a number that is the minimum
length of the coauthorship path that links them with Erd\H{o}s,
assuming there is such a path. More generally, considering a
main author, the collaborative distance of other authors from
the main author can be considered, or indeed between any
arbitrary pair of published authors. Authors who have written
publications with coauthors of Erd\H{o}s (the main author) but
not with Erd\H{o}s himself have an Erd\H{o}s number of 2. This
process can be continued in an iterative manner, using a path of
minimum length to determine the Erd\H{o}s number when there is
more than one path, as is typically the case for active
researchers in the field.

%

\begin{figure}[bth]
\begin{center}
\includegraphics[width=0.8\textwidth]{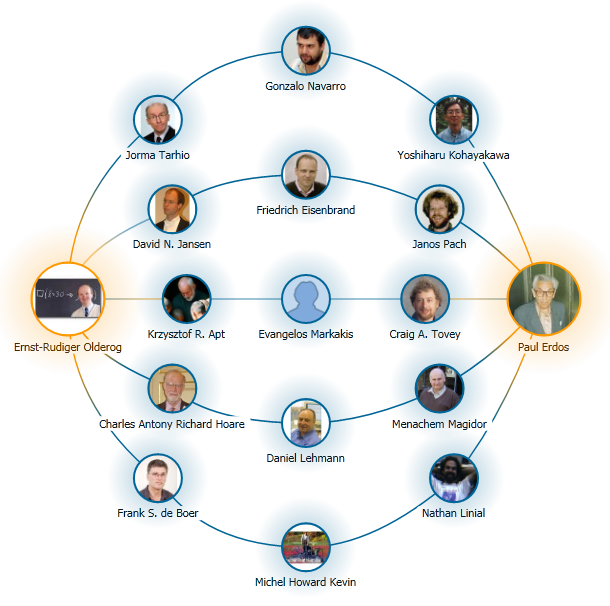}
\end{center}
\caption{A selection of connections with Paul Erd\H{o}s
for Ernst-R\"udiger Olderog on Academic Search.\label{ERO-erdos}}
\end{figure}

Academic Search can provide a graphical view of a number of the
shortest paths between any two coauthors, with Paul Erd\H{o}s
provided as the standard second author unless a different author
is explicitly selected. Figure~\ref{ERO-erdos} shows an example
for Ernst-R\"udiger Olderog. Here, five paths with a
collaborative distance of four are shown. The five researchers
on the right directly connected to Erd\H{o}s have an Erd\H{o}s
number of 1. Of the five researchers directly connected to
Olderog on the left, one (C.\ A.\ R.\ Hoare) was also on the
ProCoS project. Of course the database of authors and
publications may not be complete or accurate (e.g., especially
for authors with common names) and there could be shorter paths
between two authors in practice.

The $main$ author, as introduced earlier, could be considered as a
coordinator of a Community of Practice \cite{Wen98}. Direct
coauthors with the main coordinator take on a major editorial
role in the CoP. Those that are related to the main author by
transitive coauthorship are active members. These people form
the core of the CoP membership. Those that cite any of the above
are peripheral members of the CoP. Finally, other unrelated
published authors are considered to be outsiders to the CoP, but
are potential members.

\begin{schema}{CoP}
Researchers \\
editorial, active, core, peripheral, cop, outsiders : \finset \PEOPLE
\where
editorial = coauthors \limg \{main\} \rimg
\also
active = related \limg \{main\} \rimg \setminus editorial
\also
core = \{main\} \cup editorial \cup active
\also
peripheral = citing\_authors \limg core \rimg \setminus core
\also
cop = core \cup peripheral
\also
outsiders = published \setminus cop
\end{schema}

In the context of the ProCoS example based on E.-R.\ Olderog as
the main author. Those related by transitive authorship could be
considered core members. The collaborative distance could be
limited to some set maximum if desired. Authors that have cited
core ProCoS researchers are peripheral members of the ProCoS
community. All other published researchers are considered
outsiders to the community. Of course this formalization could be
varied if desired, but it gives a precise definition for an
informal social science concept of a CoP.

\section{Citation Metrics}
\label{metrics}

In the previous section we considered published authors and
their communities of researchers. Here we consider individual
authors and their publications.  Nowadays there are various
web-based facilities that index academic publications online,
including facilities that allow citation data to be calculated
automatically. For example, Google has a specific search
facility for indexing scholarly publications through {\em Google
Scholar} (\url{http://scholar.google.com}). Books are also
available online through {\em Google Books}
(\url{http://books.google.com}), although this does not record
citation information. Google Scholar has very complete and
up-to-date information compared to other sources \cite{Bre12},
although this can be less reliable due to the lack of human
checking. However, Google Scholar provides a facility for
individuals to generated a personalized and publicly available
web page presenting their own publications with citation
information that can be hand-corrected by the author involved as
needed at any time.

The automated search through crawling of websites including
publications with references that is undertaken by Google
Scholar is fairly reliable for publications with a reasonable
number of citations. The various citations allows automated
improvement of the information. Typically for a given author on
their personalized page, the publications list includes a ``long
tail'' of uncited or lesser cited publications, some of which
can be spurious and with poor default information. These can be
edited or deleted as required. In addition to valid
publications, Google also trawls online programme committee data
for conferences, In these cases all the committee members are
normally considered to be authors by Google Scholar.

National governments and other institutions are increasingly
keen on measuring research output by academics, with potentially
significant implications on funding for universities. For
example in the United Kingdom, the Research Excellence Framework
(REF) in 2014 and next to be held in 2020 -- formerly the
Research Assessment Exercise (RAE) until 2008 -- assesses all
research-active academics that UK universities wish to return in
various Units of Assessment (UoAs) covering all the standard
academic disciplines. There was a move to use Google Scholar
for REF 2014 Sub-panel 11 (computer science), but in the end
this we not possible for commercial reasons \cite{REF14}.

The UK REF exercise is used to allocate limited general
research funding to universities. Up to four ``best'' papers are
selected by individual academics from the period in question
(most recently six years) for assessment, normally assessed by
peer review. These should be in highly rated journals present
significant novel research ideally. The number of citations can
be a very important factor in determining the quality of a
specific paper since it provides an indication of its influence
in the field. Of course recent papers may not have had time to
receive a significant number of citations, even if they later
prove to be influential. A better indication could be obtained
by considering citations to papers in the previous assessment
period, but this is not undertaken in the UK REF at least.

There are various possible ways to measure the influence of a
researcher through their publications. One of the simplest is
the number of citations. This can vary widely between
disciplines, and of course depends on the length of the career
so far for a researcher, as well as patterns of collaboration
with other researchers. Joint publications mean that a
researcher can appear much more productive than if only
single-author publications are produced. Thus the sciences where
multi-authored papers are the norm fair better for citation
counts than the humanities where single-author books on research
are more normal. However within a given discipline (e.g.,
computer science), comparison using citation metrics has some
validity.

The total number of citations can be deceptive for reasons
dependent on the field. For researchers with a reasonable
number of publications, there is a standard pattern to the
distribution of citations for individual publications
\cite{Bre14}. Normally a researcher has a small number of
publications with significant numbers of citations (and thus
influence). Conversely there is typically a much larger number
of publications with only a few citations (and hence much less
influence). In practice the small number of highly cited
publications are much more important in terms of influence than
the larger number of lesser-cited publications. Yet the
total number of citations for the latter may be significant in
size compared with the former.

To overcome these issues, further citations metrics than just
citation counts have been developed. One of the most popular is
the {\em h-index} \cite{Hir05}. This measures the number $h$ of
publications by an individual author that have $h$ or more
citations. This provides a reasonably simple measure of the
influence of an author through their most highly cited
publications. All other lesser-cited publications have no
influence on this metric. Google Scholar includes this metric on
personal pages generated by individual researchers
automatically,

The h-index can be formalized using the Z notation
\cite{Bow01,Spi01}, for example. This was done in a functional
style in an earlier paper \cite{Bow13b}. Here we present a more
relational and arguably more abstract definition. As in the
previous paper, we use a Z ``bag'' (sometimes also called a
multiset) to model the citation count for each individual
publication. We use a generic definition for flexibility.

\def\hindex{\keyword{h-index}}

\begin{gendef}[X]
\hindex : \bag X \fun \nat
\where
\forall b: \bag X ; h:\nat \spot 
\hindex b = h \iff 
h = \# \{ x:X \mid b(x) \geq h \}
\end{gendef}

\noindent
Note that Z bags are defined as $\bag X == X \pfun \nat_1$, a
partial function from any generic set $X$ to non-zero natural
numbers. $X$ can be used to represent cited publications, for
example, mapped to the number of citations associated with each
of these publications. A publication with no citations will not
be covered in this mapping,

The h-index metric should be treated with some caution since
comparison across different academic disciplines may not be
valid due to differences in patterns of publication. In
humanities, single-author publications are the norm, as
previously mentioned. In computer science, a small number of
coauthors is typical (e.g., two to three on average), with
acknowledgements to others that have helped with the research in
some smaller way. A supervisor may be named as second author to
publication by a doctoral student, whereas in humanities the
supervisor may well not be named. In chemistry, a larger number
of coauthors is typical, with a team of people (e.g., ten or
more) working on a problem, providing different expertise.
Indeed, coauthors may not have been involved in writing the
paper at all, but may have given help with an experiment, for
example. In physics, very large numbers of coauthors are
possible for sizable and expensive initiatives (perhaps even
hundreds, e.g., experiments at CERN).

On an individual author's personalized Google Scholar page, as
set up by and editable the author, the number of citations for
each publication and the total sum of citations together with
the author's h-index and also {\em i10-index} (the number of
publications with 10 or more citations \cite{Bow13b}), are
displayed, for the last six years and for all time. A particular
aspect that is lacking in Google Scholar is any significant
visualization facility. The only visual output provided is in
the form of bar charts of the number of citations each year for
authors and also for individual papers. This is useful but not
very impressive.

As an alternative to Google Scholar, Microsoft Research's {\em
Academic Search} (see
\url{http://academic.research.microsoft.com}) provides another
online data\-base of academic publications. This was initiated
at the Microsoft Beijing research laboratory in China.
Unfortunately the resource is by no means as complete or up to
date as the information provided by Google Scholar, although
historical coverage of journals in the sciences is good. It
appears that regular updates ceased in 2012. On the positive
side, Academic Search does provide much better visualization
facilities compared to Google Scholar, as illustrated in
Section~\ref{community}. It has also been possible for any
individual to submit corrections regarding any publication entry
within the database. These have been checked by a human before
being accepted (after some variable delay).

In addition to the h-index, Academic Search also provides the
``{\em g-index}'' \cite{Egg06} for each author. This is a
refinement of the h-index and arguably provides a somewhat
improved indication of an author's academic influence.
The g-index measure gives very highly cited publications (e.g.,
a significant book or foundational paper) more weight than with
the h-index, where additional citations over and above the
h-index itself for individual publications have no effect on its
value. In the case of g-index, the most cited $g$ papers must
have at least $g^2$ citations in combination. Thus very highly
cited publications do contribute additional weight to the
g-index. Indeed, the value of the g-index is always at least as
great as the h-index for a given author and is greater if there
are some very highly cited publications.

In \cite{Bow13b}, the g-index was formally defined in Z using a
functional style, close to how its calculation could be
implemented. Here we use a more relational style of
specification, arguably more abstract and certainly less easily
directly implemented in an imperative programming language:

\def\gindex{\keyword{g-index}}
\let\sum\Sigma
\let\gsum\sum


\begin{gendef}[X]
\gindex : \bag X \fun \nat
\where
\forall b: \bag X ; g:\nat \spot \\
\t1 g*g \leq
  max \{ a:\bag X \mid a\subseteq b \land \#a=g \spot \gsum a \}
    < (g+1)*(g+1)
\end{gendef}

\noindent
Note that the $\sum$ function calculates the sum of all items in
a bag and was defined formally in \cite{Bow13b}.

%

Other citation indices include the {\em i10-index} as used on
Google Scholar, indicating the number of publications with ten
or more citations \cite{Bow13b} and the lesser used {\em f-index}
\cite{Kat09}, designed to be fairer in determining researchers
with influence across more communities. With a plethora of
citation indices, caution should be taken as to their
reliability in practice. Encouraging the production of more
papers with incremental results can be detrimental to the
advancement of scientific knowledge \cite{Par07}.

\section{Conclusion}
\label{conclusion}

This paper has presented the collaborative European ESPRIT
ProCoS projects and Working Group on Provably Correct Systems of
the 1990s and the community that this formed. It considers the
framework of a Community of Practice (CoP) in the context of
collaboration and influence within such a community through
coauthorship. We also have also considered citations to
individual publications for a particular author. The development
of knowledge depends on such communities of researchers, which
are created and then transmogrify as needed, depending on the
interests of individual researchers interacting in the larger
community.

A case study of an individual involved with the ProCoS project
has been included with visualization of connections between
researchers. Key concepts have been formalized using the Z
notation. Further formalizations and considerations of
sociological issues within the CoP framework could be considered
in more detail in the future.

As well as communities of researchers, this paper has also
discussed citation metrics for individual researchers, which
have become increasingly widespread. It should be noted that the
relevance of these, like most metrics, is a matter of debate and
any such measurements should always be treated with caution and
interpreted in an appropriate manner. In particular, the
citations at any particular point in time are a snapshot with no
precise indication of future citations. In addition, general
concepts are often not cited as all. Many disciplines have a
practice of including ``passive'' authors that have not directly
undertaken the research, perhaps acting as a supervisor or
funder instead. These and other issues mean that all citation
statistics should be used with caution.

Possible future directions include considering the graphs of
relationships between authors and publications more holistically
to model movements and influences, but this is beyond the scope
of this paper.

\paragraph*{Acknowledgements:}
Jonathan Bowen is grateful for financial support from Museophile
Limited. Thank you to Microsoft for the Academic Search
facility, which provided the screenshots for the figures in this
paper. Many thanks to collaborators on the ProCoS projects and
Working Group during the 1990s and subsequently. A special
thank you to Prof.\ Dr Ernst-R\"udiger Olderog of Oldenburg
University for the individual case study. The Z notation in
this paper has been type-checked using the \fuzz\ type-checker
\cite{Spi08}. Finally, the reviewers provided helpful comments
that improved the presentation and content of the paper.

\bibliographystyle{splncs03}
\bibliography{ero2015,procos2,procos-wg}

\end{document}